\documentclass[aps,prl,floatfix,superscriptaddress,reprint]{revtex4-2}
\usepackage{mathrsfs,braket}
\usepackage{amssymb, amsbsy, amsmath, latexsym, dsfont, array, layout, graphicx, mathrsfs, color, ulem, bm}
\usepackage[colorlinks,linkcolor=blue,anchorcolor=red,citecolor=blue]{hyperref}

\begin{document}
\title{Anderson Transition and Mobility Edges in a Family of 3D Fractal Lattices}

\author{Tianyu Li}
\email{tianyuli@m.scnu.edu.cn}
\affiliation {Key Laboratory of Atomic and Subatomic Structure and Quantum Control (Ministry of Education), Guangdong Basic Research Center of Excellence for Structure and Fundamental Interactions of Matter, School of Physics, South China Normal University, Guangzhou 510006, China}
\affiliation {Guangdong Provincial Key Laboratory of Quantum Engineering and Quantum Materials, Guangdong-Hong Kong Joint Laboratory of Quantum Matter, Frontier Research Institute for Physics, South China Normal University, Guangzhou  510006, China}
\author{Xin Tang}
\affiliation {Key Laboratory of Atomic and Subatomic Structure and Quantum Control (Ministry of Education), Guangdong Basic Research Center of Excellence for Structure and Fundamental Interactions of Matter, School of Physics, South China Normal University, Guangzhou 510006, China}
\affiliation {Guangdong Provincial Key Laboratory of Quantum Engineering and Quantum Materials, Guangdong-Hong Kong Joint Laboratory of Quantum Matter, Frontier Research Institute for Physics, South China Normal University, Guangzhou  510006, China}
\author{Sheng Liu}\email{shengliu@ustc.edu.cn}
\affiliation{Key Laboratory of Quantum Information, University of Science and Technology of China, Hefei 230026, China}
\author{Haiping Hu}
\email{hhu@iphy.ac.cn}
\affiliation{Beijing National Laboratory for Condensed Matter Physics, Institute of Physics, Chinese Academy of Sciences, 100190 Beijing, China}
\affiliation{School of Physical Sciences, University of Chinese Academy of Sciences, 100049 Beijing, China}

\begin{abstract}

Anderson localization is fundamentally controlled by dimensionality, yet the nature of the Anderson transition in continuously tunable noninteger dimensions remains largely unexplored. Here, we introduce a family of three-dimensional fractal lattices with continuously tunable spectral dimension $d_s\in[2,3]$, providing a controlled platform for studying localization physics beyond integer dimensions and across the lower critical dimension $d_s=2$. 
Using large-scale finite-size scaling analysis, we systematically investigate the Anderson transition and identify mobility edges throughout the fractal family. The critical disorder strength evolves continuously from $0$ to $16.6$ as the spectral dimension increases from $2$ to $3$. We show that the spectral dimension predominantly governs the universality class of the transition, while the precise critical point is additionally influenced by microscopic geometric details of the underlying fractal lattice. The critical exponent exhibits an approximate inverse dependence on $d_s$, providing quantitative insight into scaling theory in noninteger dimensions. 
Our results establish tunable fractal lattices as a versatile framework for exploring localization and quantum critical phenomena beyond conventional integer-dimensional systems.
\end{abstract}

\maketitle

{\it Introduction.---}
Anderson localization (AL), a single-particle disorder-induced phenomenon leading to the exponential localization of wave functions and the absence of wave diffusion in disordered media~\cite{anderson, Lee1985, Kramer1993, Evers2008, Lagendijk2009,thouless1973,anderson1979}, is a cornerstone of condensed matter physics. The dimensionality of the system plays a crucial role in localization phenomena. In conventional Euclidean lattices with integer spatial dimensions, it is well established that arbitrarily weak disorder induces localization in one and two dimensions, whereas in three dimensions a finite disorder strength is required to drive the Anderson transition\cite{anderson1979}.
In three-dimensional systems, the presence of a mobility edge leads to an energy-dependent localization transition, where extended states near the band center coexist with localized states near the spectral edges~\cite{Evers2008}. 
Experimentally, Anderson localization has been observed across a wide range of platforms, including ultracold atoms~\cite{exp1,exp2,exp3}, microwaves~\cite{exp4,exp5}, and sound waves~\cite{exp6}.

Anderson transition has been explored not only in high dimensions~\cite{prb2007} and effectively infinite-dimensional systems such as the Bethe lattice and random graphs~\cite{anderson73,zirnbauer,fyodorov,Altshuler1,Altshuler2,mirlin16,mirlin162,mirlin19prb,mirlin21,tarzia,tarzia2,parisi,sierant,Garel,prb2022}, and even beyond Euclidean geometries, for example in hyperbolic lattices~\cite{hyper1,hyper2}.
In contrast, much less is known about localization in systems with non-integer dimensions\cite{prl1996,prb2017}. Fractal lattices, which exhibit non-integer dimensions, therefore provide a unique platform for exploring localization beyond conventional integer-dimensional systems. 
In such systems, several notions of dimension must be distinguished: the embedding Euclidean dimension $D$, the Hausdorff dimension $d_H$, and the spectral dimension $d_S$\cite{prl1996,prb2017}. While $d_H$ characterizes the geometric scaling of the lattice, $d_S$ is associated with random-walk dynamics and governs diffusion and transport properties.  Importantly, Anderson transition is primarily governed by the spectral dimension $d_S$~\cite{prl1996,prb2017,pre1994}, rather than $D$ or $d_H$, which coincide in integer-dimensional lattices.

In this work, we construct a family of fractal lattices embedded in three-dimensional Euclidean space, whose spectral dimension $d_S$ can be continuously tuned over the range $d_S\in[2,3]$ through controllable growth rules and bond-density modifications. Unlike previous studies of Anderson localization on fractals, which were limited to a few isolated lattice geometries with fixed spectral dimensions~\cite{prl1996}, our framework provides a unified platform for systematically exploring localization physics across the lower critical dimension $d_S=2$.
This tunability enables us to investigate how dimensionality controls Anderson criticality in noninteger-dimensional systems. By analyzing the mobility edge and performing finite-size scaling of the level-spacing ratio, we extract the critical disorder strength and localization-length exponent throughout the continuously tunable parameter regime. We find that the spectral dimension predominantly determines the universality class of the transition, while the precise critical disorder depends sensitively on microscopic geometric details of the fractal lattice. 
Furthermore, we uncover an approximate inverse relation between the critical exponent and spectral dimension,
\begin{equation}
\nu \approx \frac{4.29}{d_S}+0.08,
\end{equation}
which remains valid across the broad interval $2\lesssim d_S\lesssim3$. This scaling relation provides quantitative evidence for dimensional control of Anderson criticality in fractal systems and may serve as a useful benchmark for renormalization-group descriptions in noninteger dimensions.
Our work establishes tunable fractal lattices as a versatile platform for investigating localization phenomena beyond conventional integer-dimensional systems.

%\date{\today}

{\it 3D fractal Model.---}
We study Anderson localization on three-dimensional fractal lattices generated by dielectric breakdown\cite{2Dfractal}. The system consists of a dielectric medium confined to a sphere with electrostatic potential fixed to $V=1$ at the surface and $V=0$ at the center. The volume is discretized on a cubic lattice with approximately $190$ sites along the diameter. Breakdown initiates at the center and proceeds by the successive failure of nearest-neighbor bonds connecting the discharged cluster to the surrounding lattice.

At each growth step, The electric potential $V_{ijk}$ of the undischarged lattices is obtained by solving the discrete Laplace equation \cite{2Dfractal,3Dfractal}
\begin{equation}
V_{ijk}=\frac{1}{6}\sum_{\langle i'j'k'\rangle} V_{i'j'k'}, \label{eqV}
\end{equation}
where the sum runs over the six nearest neighbors of site $(i,j,k)$. This corresponds to the discretized form of the Laplace equation, $\nabla^{2} V = 0$, on the cubic lattice. The discharged cluster forms an equipotential region with $V=0$ at all stages.  Equation~(\ref{eqV}) is solved iteratively under the boundary conditions that the potential is fixed at $V=1$ on the spherical boundary and $V = 0$ on the discharged cluster. The potential at all other sites is initially set to $V = 1/2$.  The iteration of Eq.~(\ref{eqV}) is then performed for $30\sim 200$ times, after which the potential converges to a solution that satisfies Eq.~(\ref{eqV}).

\begin{figure}[h]
\centering
\includegraphics[width=0.5\textwidth]{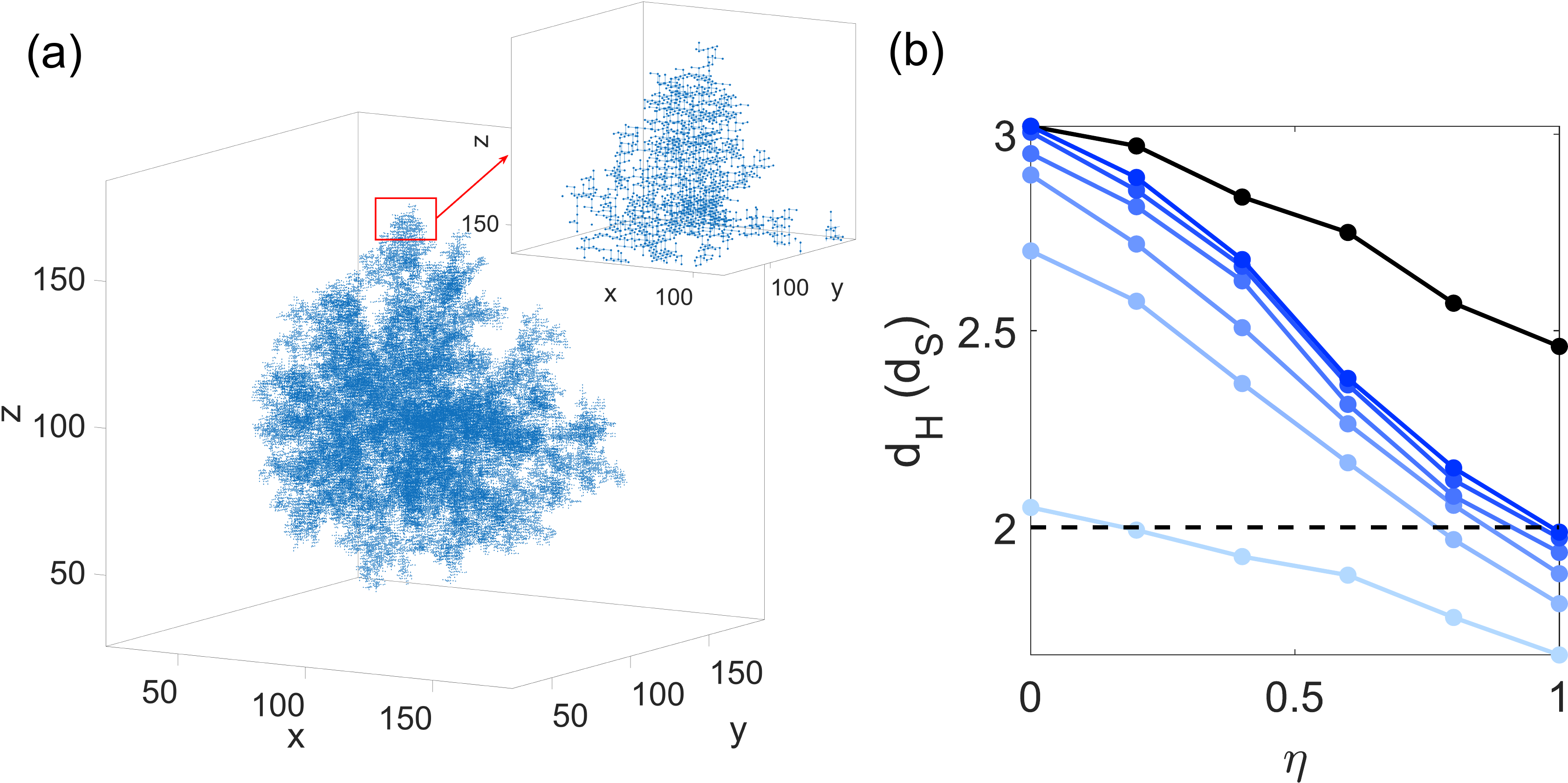}
\caption{(a) Fractal structure for $\eta = 1$.
(b) Hausdorff dimension $d_H$ and spectral dimension $d_S$ as functions of the parameter $\eta$. The black line denotes the Hausdorff dimension $d_H$, which is independent of $p$. The blue curves represent the spectral dimension $d_S$ for $p = 0, 0.2, 0.4, 0.6, 0.8,$ and $1$, shown with a gradient of blue shades from light to dark. The system size are taken as $N=80000$.}
\label{fig1}
\end{figure}

With the potential field, for each candidate bond connecting a discharged site $(i,j,k)$ to an undischarged neighbor $(i',j',k')$, a single bond breaks with probability proportional to the local electric field raised to a power $\eta$, the growth probability is assigned as
\begin{equation}
p(i,j,k \rightarrow i',j',k') =
\frac{\phi_{i',j',k'}^{\eta}}{\sum \phi_{i',j',k'}^{\eta}},
\end{equation}
where $\phi_{i',j',k'}$ denotes the local potential, and the summation runs over all candidate bonds. After each breakdown event, the Laplace equation is solved iteratively to update the potential field. The process is terminated after approximately $\sim 10^5$ discharged bonds.  This construction corresponds to the three-dimensional extension of the Niemeyer-Pietronero-Wiesmann (NPW) model\cite{2Dfractal,3Dfractal}. An illustrative example of the resulting fractal lattice for $\eta =1$ is shown in Fig.~\ref{fig1}(a). The discharged cluster forms a single connected, loopless graph (i.e., a tree) with many branches. These branches exhibit self-similarity and characteristic fractal behavior. In this structure, the removal of any bond disconnects the lattice, we refer to this as the minimal graph.

By subsequently adding bonds without introducing new discharge sites, one can construct a maximal graph. At this stage, all possible bonds that do not create new discharge sites have already been added, and any further bond addition necessarily generates new discharge sites~\cite{prb2017}. Intermediate structures are parametrized by a probability $p$, defined as the fraction of randomly added bonds, with $p=0$ ($p=1$) corresponding to the minimal (maximal) graph. For convenience, we use $(\eta,p)$ to label different lattices.

The geometry of the resulting lattices is characterized by the Hausdorff dimension $d_H$, defined via $N(r)\sim r^{d_H}$, where $N(r)$ is the number of sites within a sphere of radius $r$. While the transport properties are governed by the spectral dimension $d_S$, which we extract from random-walk simulations\cite{prl1996,pre1994}. Specifically, we simulate a random walker on the fractal lattice and analyze the scaling of the mean-square displacement $\langle r^2(t)\rangle$ with time $t$. On fractal lattices, diffusion is anomalous, scale as 
$\langle r^2(t)\rangle \sim t^{\alpha},\alpha < 1,$
in contrast to normal diffusion in regular lattices where $\alpha = 1$. The exponent $\alpha$ is related to the Hausdorff dimension through
$
d_S = \alpha d_H
$
\cite{pre1994}, implying $d_S < d_H$ and reflecting reduced transport efficiency on fractal geometries.

By tuning $\eta$ and $p$, both the spectral dimension $d_S$ and the Hausdorff dimension $d_H$ can be continuously varied. As shown in Fig.~\ref{fig1}(b), $d_H$ is independent of $p$ and decreases with increasing $\eta$, ranging from $2.46$ to $3$. In contrast, $d_S$ depends strongly on both parameters and can be tuned over a broad range from $1.68$ to $3$. In particular, increasing $p$ enhances the connectivity of the lattice, leading to a larger spectral dimension while keep $d_H$ unchanged.

In the limit $\eta \to 0$, $d_H$ approaches $3$, corresponding to a regular three-dimensional  cubic lattice.
 For $\eta = 1$, we obtain $d_H = 2.46$, consistent with the value $d_H = 2.48$ reported in Ref.~\cite{3Dfractal}. 
Interestingly, for the minimal graph of the three-dimensional  cubic lattice (with $(\eta,p)=(0,0)$), although $d_H = 3$, the spectral dimension approaches $d_S \approx 2$, suggesting the absence of Anderson localization in this case, as confirmed by our numerical results (see Supplemental Material).
Moreover, we find numerically that for $p = 0$, the relation $d_S \simeq \frac{2}{3} d_H$ holds for all values of $\eta$.

The spectral dimension is closely related to Anderson localization, which occurs only for $d_S > 2$ \cite{anderson1979}. In the constructed lattices, we find that $2 < d_S < 3$ over a broad range of $\eta$ and $p$, placing the system above the lower critical dimension for Anderson localization and thereby allowing an Anderson transition to occur in these fractal lattices.

{\it Mobility edges on 3D fractals.---}
We now investigate the effect of disorder on the fractal lattice described by the Hamiltonian
\begin{align}
H=\sum_{\langle i,j\rangle}t(c_i^\dagger c_{j} + \mathrm{h.c.})+\sum_j \epsilon_j c_j^\dagger c_j.
\end{align}
Here, $c_j^\dagger$ ($c_j$) denotes the creation (annihilation) operator at site $j$. The hopping amplitude $t$ is set as the energy unit ($t=1$). The onsite disorder $\epsilon_j$ is uniformly distributed in the interval $[-W/2, W/2]$.

To characterize the Anderson transition on the fractal lattice, we employ two quantities. The first is the ratio of adjacent level spacings, defined as
\begin{align}
r_i = \frac{\min(\delta_i, \delta_{i-1})}{\max(\delta_i, \delta_{i-1})},
\end{align}
where $\delta_i = E_i - E_{i-1}$ and $\{E_i\}$ are the eigenenergies arranged in ascending order. 
In the localized phase, the level statistics follow a Poisson distribution with $\langle r \rangle \approx 0.387$, whereas in the delocalized phase they follow a Wigner--Dyson distribution with $\langle r \rangle \approx 0.529$. Here, $\langle r \rangle$ is averaged over eigenstates and disorder realizations.

To further characterize the spatial structure of eigenstates, we introduce the fractal dimension ($D_2$) of a normalized eigenstate,
\begin{equation}
D_2 = -\frac{\ln \sum_n |\psi_n|^4}{\ln N},
\label{eq:FD}
\end{equation}
where $\psi_n$ denotes the amplitude at site $n$, and $N$ is the system size. In the thermodynamic limit, the FD approaches 0 for localized states and 1 for extended states.

\begin{figure}[h]
\centering
\includegraphics[width=0.5\textwidth]{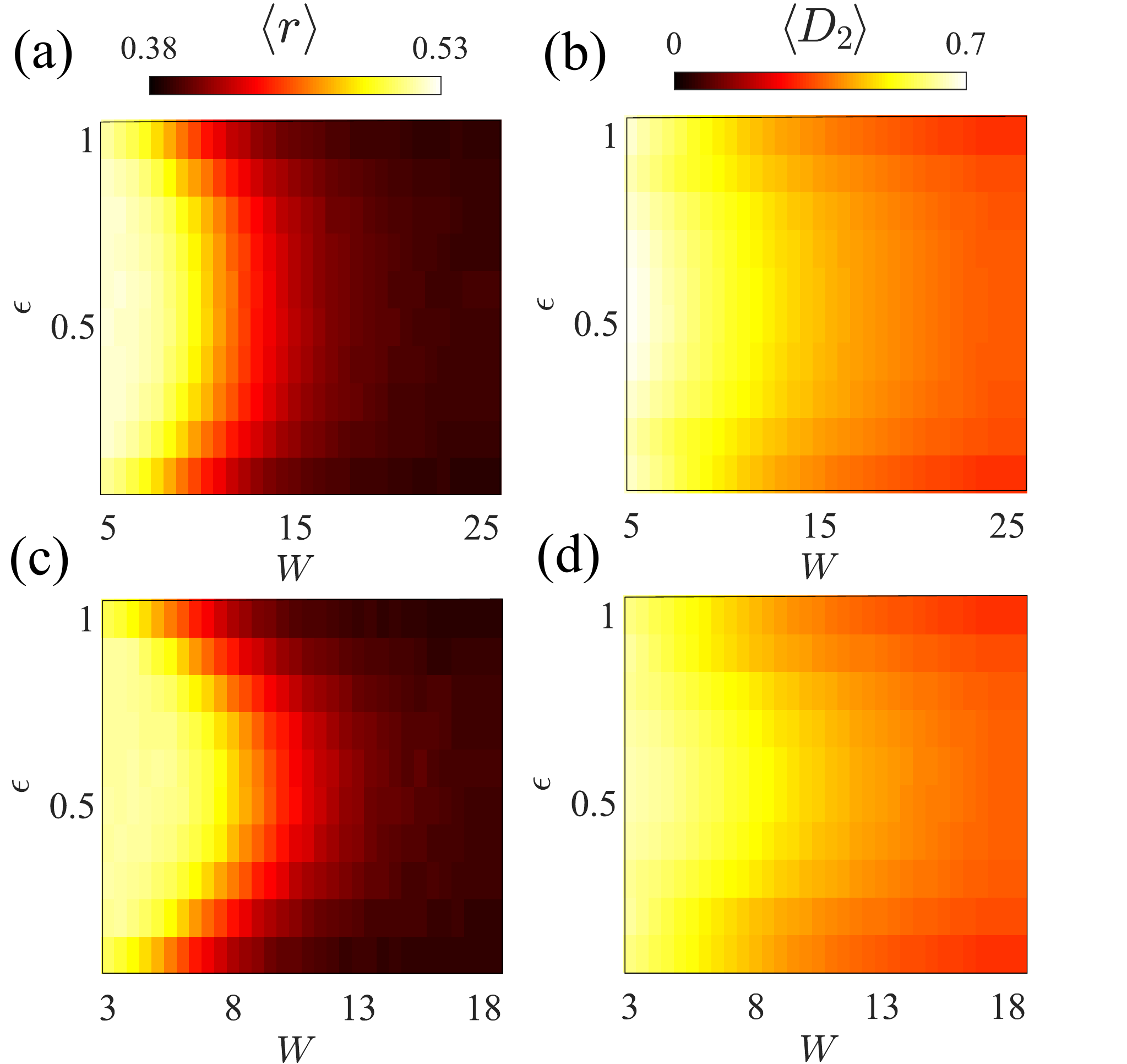}
\caption{ The mobility edges. The presence of mobility edges in the energy spectra of the model with $(\eta,p) = (0.2,1)$ (a),(b) and $(\eta,p) = (0,0.4)$ (c),(d). (a)(c) The average ratio of level spacings $\langle r\rangle$ and (b)(d) average Fractal demension $\langle D_2\rangle$ with respect to the disorder strength and eigenenergies.  The eigenenergies are divided into ten equal segments and rescaled to the interval $\epsilon\in[0,1]$. The averages are taken within each segment and over $200$ disorder realizations. The system size are taken as $N=5000$.}
\label{fig2}
\end{figure}

To identify the emergence of mobility edges (MEs) with increasing disorder strength, we partition the energy spectrum into ten segments, each containing an equal number of eigenstates, and analyze the properties of eigenstates within different energy windows. We introduce a rescaled energy variable $\epsilon = C(E)/N$, where $C(E)$ denotes the number of eigenstates with energy below $E$, and $N$ is the total number of states, such that $\epsilon \in [0,1]$.
By averaging over eigenstates within each energy segment as well as over $200$ disorder realizations, we compute the level-spacing ratio $\langle r \rangle$ and the fractal dimension $\langle D_2 \rangle$ across the spectrum. As shown in Fig.~\ref{fig2}, the eigenenergies are divided into ten equal segments and rescaled to the interval $\epsilon \in [0,1]$, allowing us to resolve the energy dependence of localization properties.

In Fig.~\ref{fig2}(a) and \ref{fig2}(b), we present the results for $(\eta,p) = (0.2,1)$. The transition between extended and localized behavior exhibits a strong energy dependence. States near the band center (intermediate $\epsilon$) remain extended up to relatively large disorder strengths, as indicated by $\langle r \rangle \approx 0.53$ and finite $\langle D_2 \rangle$. In contrast, states near the spectral edges ($\epsilon \to 0$ or $1$) localize at much weaker disorder, characterized by $\langle r \rangle \approx 0.39$ and vanishing $\langle D_2 \rangle$. This pronounced energy dependence provides clear evidence for the existence of mobility edges in the fractal lattice.
Similarly, Figs.~\ref{fig2}(c) and \ref{fig2}(d) show results for $(\eta,p) = (0,0.4)$, displaying the same qualitative behavior. In both cases, the contour plots of $\langle r \rangle$ and $\langle D_2 \rangle$ consistently reveal a boundary in the energy--disorder plane separating extended and localized regimes. The agreement between spectral statistics and wavefunction fractality further supports the identification of mobility edges.
The observed behavior of the mobility edges is qualitatively similar to that in the three-dimensional Anderson model~\cite{Evers2008} and in hyperbolic lattices~\cite{hyper1}, indicating that fractal lattices with tunable spectral dimension provide a natural extension of localization physics beyond conventional integer-dimensional systems.

\begin{figure}[h]
\centering
\includegraphics[width=0.5\textwidth]{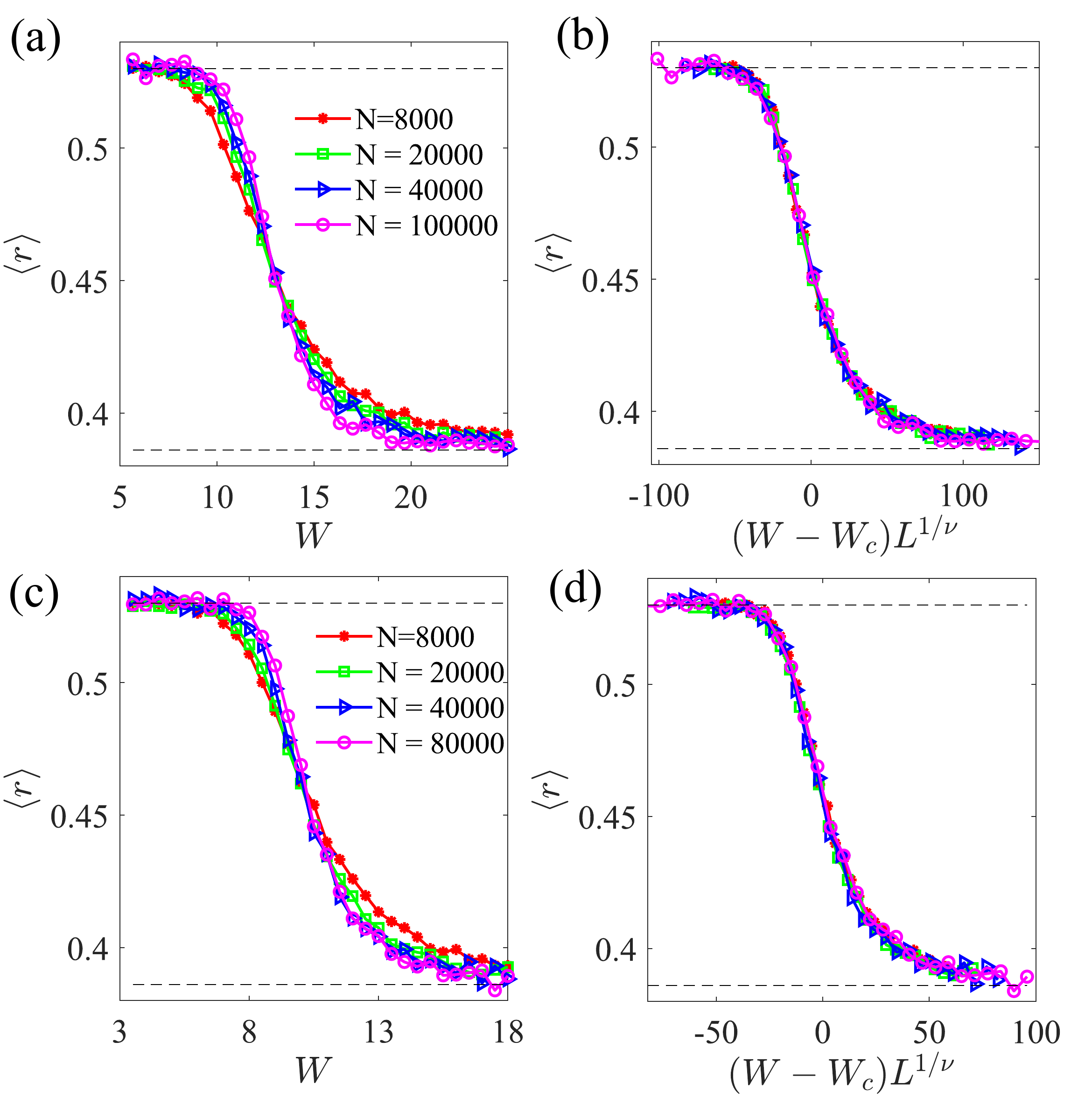}
\caption{Finite-size scaling of the level statistics. The average ratio of level spacings $\langle r\rangle$ as a function of disorder strength $W$ for different system sizes: (a) $(\eta ,p) = (0.2,1)$ and (c) $(\eta ,p) = (0,0.4)$. Panels (b) and (d) show the corresponding finite-size scaling collapses of (a) and (c), respectively, with the scaling variable $L =N^{1/d_S}$, where $N$ is the number of sites.
For (a) and (b), the extracted critical disorder strength and correlation-length exponent are $W_c = 12.9\pm0.1$ and $\nu = 1.54\pm0.04$, with spectral dimension $d_S = 2.89$. For (c) and (d), corresponding to $(\eta ,p) = (0,0.4)$, we obtain $W_c = 10.2\pm0.2$ and $\nu = 1.59\pm 0.06$, with $d_S = 2.90$.
The averages are taken over the central 10 eigenstates and $2000$–$10000$ disorder realizations. The horizontal dashed lines indicate $\langle r\rangle = 0.387$ (Poisson) and $\langle r\rangle = 0.529$ (Wigner–Dyson).
}
\label{fig3}
\end{figure}

{\it Finite-size scaling of the level statistics.—}
To accurately determine the transition point $W_c$ and extract the critical exponent $\nu$, we focus on the ten eigenvalues closest to zero (the band center) and perform a finite-size scaling analysis of the level-spacing ratio $\langle r \rangle$, averaged over both these eigenvalues and disorder realizations. In Fig.~\ref{fig3}(a), we show the dependence of $\langle r \rangle$ on the disorder strength for different system sizes of the $(\eta, p) = (0.2, 1)$ fractal lattice, with sizes up to $N = 100000$ and $2000$--$10000$ disorder realizations. The value of $\langle r \rangle$ decreases from $0.529$ in the delocalized regime at weak disorder to $0.387$ in the localized regime at strong disorder.
We then employ the finite-size scaling form $\langle r \rangle = F\big[(W - W_c)L^{1/\nu}\big]$, where $L = N^{1/d_S}$, with $N$ the number of lattice sites and $d_S$ the spectral dimension. For this lattice, $d_S = 2.89$. After rescaling, data for different system sizes collapse onto a single universal curve. From this collapse, we extract the best-fit values $W_c = 12.91\pm0.1$ and $\nu = 1.54\pm0.04$ (see Supplemental Material), as shown in Fig.~\ref{fig3}(b).
For the $(\eta, p) = (0, 0.4)$ lattice with system sizes up to $N = 80000$, we similarly obtain $W_c = 10.2\pm0.2$ and $\nu = 1.59\pm0.06$, with $d_S = 2.90$, as shown in Figs.~\ref{fig3}(c,d).
Notably, although the spectral dimensions of the $(\eta, p) = (0.2, 1)$ and $(\eta, p) = (0, 0.4)$ lattices are nearly identical, the corresponding critical disorder strengths differ by more than $20\%$. This observation indicates that the spectral dimension alone does not fully determine the transition point. Instead, the underlying lattice geometry also plays a secondary but non-negligible role in shaping the critical behavior.

\begin{figure}[h]
\centering
\includegraphics[width=0.5\textwidth]{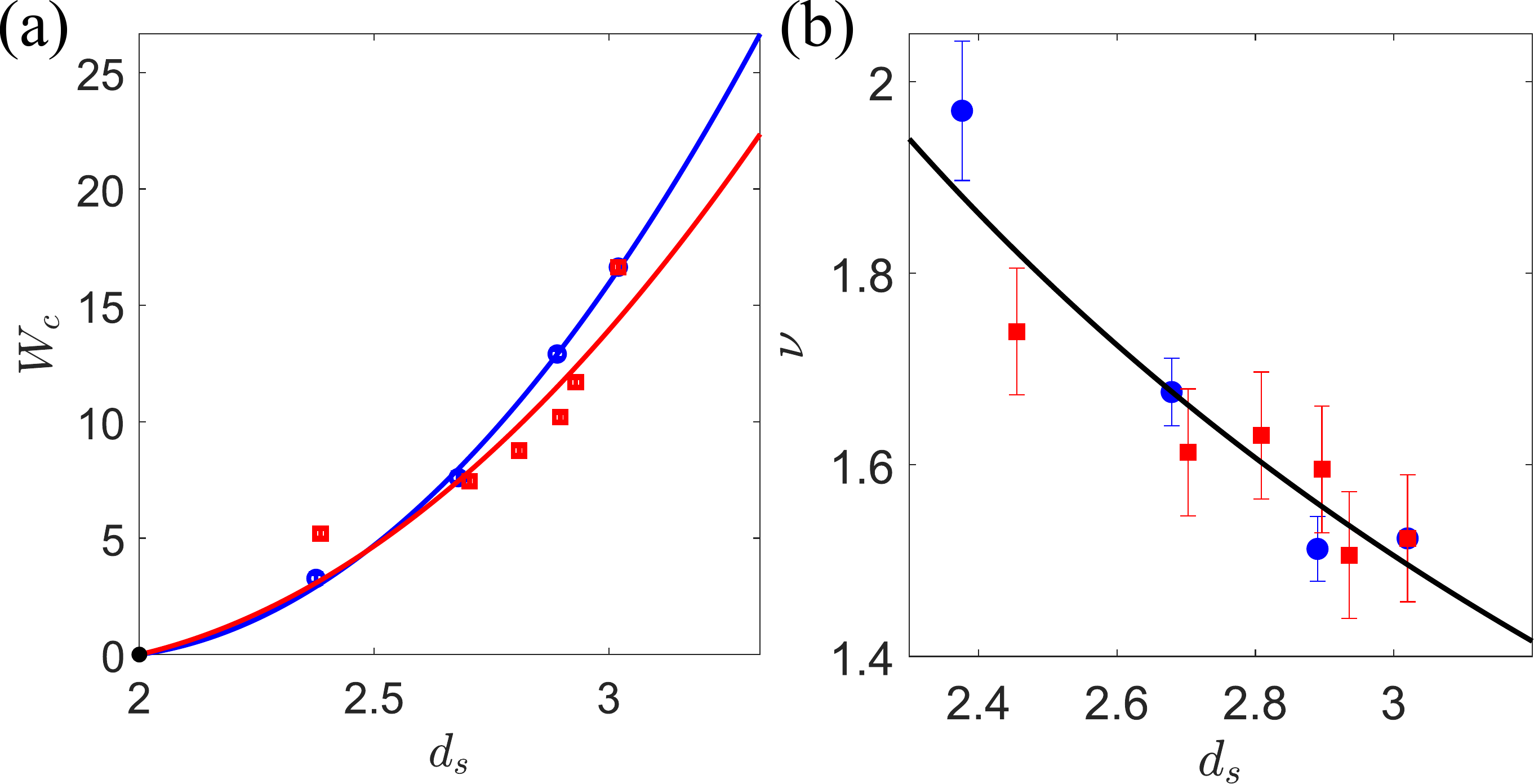}
\caption{(a) Anderson transition point $W_c$ as a function of the spectral dimension. The blue circles correspond to results obtained at fixed $p = 1$, with $\eta = 0, 0.2, 0.4,$ and $0.6$. The red squares correspond to results at fixed $\eta = 0$, with $p = 0.1, 0.2, 0.3, 0.4, 0.5,$ and $1$. The blue and red curves represent quadratic fits to the respective datasets, with $W_c = 2.83(d_S-2) + 13.11(d_S-2)^2$ and $W_c = 4.71(d_S-2) + 9.23(d_S-2)^2$ respectively.
(b) Critical exponents corresponding to panel (a). The two datasets are combined, and the black line shows the fitted result with  $\nu = 4.29/D_s + 0.08$. }
\label{fig4}
\end{figure}

To elucidate the dependence of the transition point $W_c$ and the critical exponent $\nu$ on the spectral dimension $d_S$, we systematically study two groups of fractal lattices characterized by different $(\eta, p)$ parameters. For each case, we determine $d_S$ and extract $W_c$ and $\nu$ via finite-size scaling.
In the first group, we fix $\eta = 0$ and vary $p = 0.1, 0.2, 0.3, 0.4, 0.5, 1$, corresponding to spectral dimensions $d_S = 2.38, 2.70, 2.81, 2.90, 2.93, 3.00$. In the second group, we fix $p = 1$ and vary $\eta = 0, 0.2, 0.4, 0.6$, with $d_S = 3.00, 2.89, 2.68, 2.38$. The extracted values of $W_c$ and $\nu$ are summarized in Fig.~\ref{fig4}, which allows for a direct comparison of their dependence on $d_S$.

Fig.~\ref{fig4}(a) shows the critical disorder strength $W_c$ as a function of the spectral dimension $d_S$. The blue circles correspond to lattices with fixed $p = 1$ and varying $\eta$, while the red squares represent lattices with fixed $\eta = 0$ and varying $p$. In both cases, $W_c$ exhibits a clear dependence on $d_S$ and is well described by a quadratic form. Notably, the two datasets do not collapse onto a single universal curve, indicating that $d_S$ alone is insufficient to fully characterize the transition point.
 The quadratic fits for the two datasets are given by
$W_c = 2.83(d_S - 2) + 13.11(d_S - 2)^2\quad \text{and} \quad W_c = 4.71(d_S - 2) + 9.23(d_S - 2)^2,$
respectively, by fitting, we have imposed the constrained to pass through the point $(d_S, W_c) = (2, 0)$.
These results reinforce the conclusion that while the spectral dimension $d_S$ plays a dominant role in determining the universality of the Anderson transition, the precise location of the transition point $W_c$ is also influenced by additional geometric features of the underlying fractal lattice.

We further investigate the dependence of the critical exponent $\nu$ on $d_S$. Fig.~\ref{fig4}(b) presents the corresponding values of $\nu$. By combining the two datasets and assuming a fitting form $\nu = \alpha / d_S + c$, where $\alpha$ and $c$ are fitting parameters, we obtain the best fit
\begin{equation}
\nu = \frac{4.29}{d_S} + 0.08.
\end{equation}
Remarkably, this relation remains valid over a broad range of spectral dimensions from $d_S\approx2$ to $d_S\approx3$, suggesting that the spectral dimension is the primary parameter governing Anderson criticality in fractal lattices. This behavior can be qualitatively understood from the return probability of diffusive motion on a fractal lattice, $P(t)\sim t^{-d_S/2}$ \cite{randomwalk}.
For smaller $d_S$, the return probability $P(t)$ decays more slowly with the number of walk steps $t$ due to the abundance of dead ends in the fractal geometry, which strongly enhances coherent backscattering and localization effects. Consequently, increasingly larger length scales are required to distinguish localized from delocalized behavior, making the transition substantially broader and yielding a larger critical exponent $\nu$.
In contrast, larger $d_S$ improves lattice connectivity and suppresses return processes, leading to a sharper transition and a smaller $\nu$. This physical picture naturally explains the approximate inverse relation between $\nu$ and $d_S$. Similar inverse relations between $\nu$ and effective dimensionality have also been reported in disordered and fractal systems~\cite{prl1996,prb2007,RG}, further supporting the universality of our results.

{\it Main conclusions.—}
In this work, we construct a family of three-dimensional fractal lattices generated via a dielectric breakdown model, with continuously tunable fractal dimensions ranging from $1.67$ to $3$. This framework provides a controllable platform for investigating localization and quantum critical phenomena beyond conventional integer-dimensional systems.
We study Anderson localization on these lattices and demonstrate the existence of mobility edges across a broad range of spectral dimensions. Our results show that the spectral dimension $d_S$ predominantly determines the universality class of the Anderson transition, whereas the critical disorder strength $W_c$ is additionally affected by microscopic geometric details of the underlying lattice. Furthermore, the critical exponent $\nu$ exhibits an approximate inverse dependence on $d_S$, revealing a quantitative relation between Anderson criticality and noninteger dimensionality.
Our work establishes tunable fractal lattices as a versatile setting for exploring localization physics in complex geometries. Our findings open avenues for future studies of localization in fractal geometries, such as the emergence of critical phases under quasiperiodic potentials. In addition, analytical approaches, including renormalization group methods~\cite{RG} and cavity methods~\cite{parisi}, may provide deeper insight into the transition in the thermodynamic limit, which we leave for future investigation.

\acknowledgments
This work is supported by the National Natural Science Foundation of China (Grant Nos. 12504189), and the Guangdong Basic and Applied Basic Research Foundation (Grant No.2026A1515012098). H.H. is supported by National Key Research and Development Program of China (Grants No. 2022YFA1405800 and No. 2023YFA1406704) and National Natural Science Foundation of China (Grant No. 12474496 and No. 12547107).  S.L. is supported by the Fundamental Research Funds for the Central Universities (WK2030000088).

\clearpage
\appendix
\setcounter{equation}{0}  
\setcounter{figure}{0}
\counterwithout{equation}{section}  
\renewcommand{\thefigure}{S\arabic{figure}}
\renewcommand{\thesection}{S\arabic{section}}
\pagebreak
\renewcommand{\theequation}{S\arabic{equation}}
\widetext
\begin{center}
\textbf{\large  Supplemental Material for ``Anderson Localization and Mobility Edges in Family of 3D Fractal Lattices"}
\end{center}

This Supplemental material (SM) provides additional details on (I) Finite-Size scaling methods; (II)Level statistics for $\eta = 0$; (III) Mean Fractal Dimension as a Function of Disorder Strength.

\section{(I) Finite-Size scaling methods}

We employ finite-size scaling analysis to determine the transition point $W_c$ and the critical exponent $\nu$. We define the following cost (error) function
\begin{align}
\mathcal{E}_r = \frac{1}{W_s^{\max}-W_s^{\min}}
\int_{W_s^{\min}}^{W_s^{\max}} dW_s \,
\mathrm{Var}_L \left\{ X_L\!\left(W_s L^{-1/\nu} + W_c \right) \right\},
\end{align}
where $X_L(W)$ denotes $\langle r(W,L)\rangle$, and $\mathrm{Var}_L$ represents the variance over different system sizes $L$.
Since $\langle r(W,L)\rangle$ and $\langle \mathrm{IPR}(W,L)\rangle$ are obtained at discrete values of $W$, we evaluate $X_L\!\left(W_s L^{-1/\nu} + W_c\right)$ using interpolation. By choosing appropriate ranges $W_s^{\min}$ and $W_s^{\max}$, the optimal values of $W_c$ and $\nu$ (or $\gamma$) are determined by minimizing $\mathcal{E}_r$.  
To estimate the uncertainties in $W_c$ and $\nu$, we vary the fitting window by sampling $W_s^{\min}$ and $W_s^{\max}$  within a broad range, for example $W_s^{min}\in [-100,-50]$ and  $W_s^{max}\in [50,100]$. Different choices of the fitting window lead to slightly different optimal values of $W_c$ and $\nu$, which we use to estimate the corresponding fitting errors.

\section{(II)Level statistics for $\eta = 0$}

When considering the breakdown process with $\eta = 0$, Eq.~(2) in the main text shows that all bonds break with equal probability. In this case, the discharged cluster grows isotropically in all directions, forming a minimally connected cluster embedded in a three-dimensional Euclidean lattice. The resulting graph has a Hausdorff dimension $d_H = 3$.

However, its spectral dimension is approximately $d_S \approx 2$, which implies the absence of a finite Anderson transition (i.e., no nonzero critical disorder strength). Consequently, even weak disorder is sufficient to induce localization. We verify this behavior using level statistics, as shown in Fig.~S1(a). For $(\eta,p) = (0,0)$, the average level-spacing ratio $\langle r \rangle$ is plotted as a function of disorder strength $W$ for different system sizes. The averages are taken over the central 10 eigenstates and $2000$--$10000$ disorder realizations. As seen in the figure, no clear crossing point emerges for different system sizes, and the values of $\langle r \rangle$ remain close to the Poisson limit, indicating that the system is in the localized regime for all disorder strengths.

In contrast, for the maximally connected case at $\eta = 0$, i.e., $(\eta,p) = (0,1)$, the lattice reduces to a regular three-dimensional Euclidean lattice with $d_H = d_S = 3$. In this case, a finite Anderson transition is observed at $W_c \approx 16.6$, as expected (see Fig.~S1(b)). As the probability $p$ is tuned from $0$ to $1$, the spectral dimension $d_S$ increases from $2$ to $3$, accompanied by the transition point shifting from $0$ to $16.6$.

\begin{figure}[h]
\centering
\includegraphics[width=0.8\textwidth]{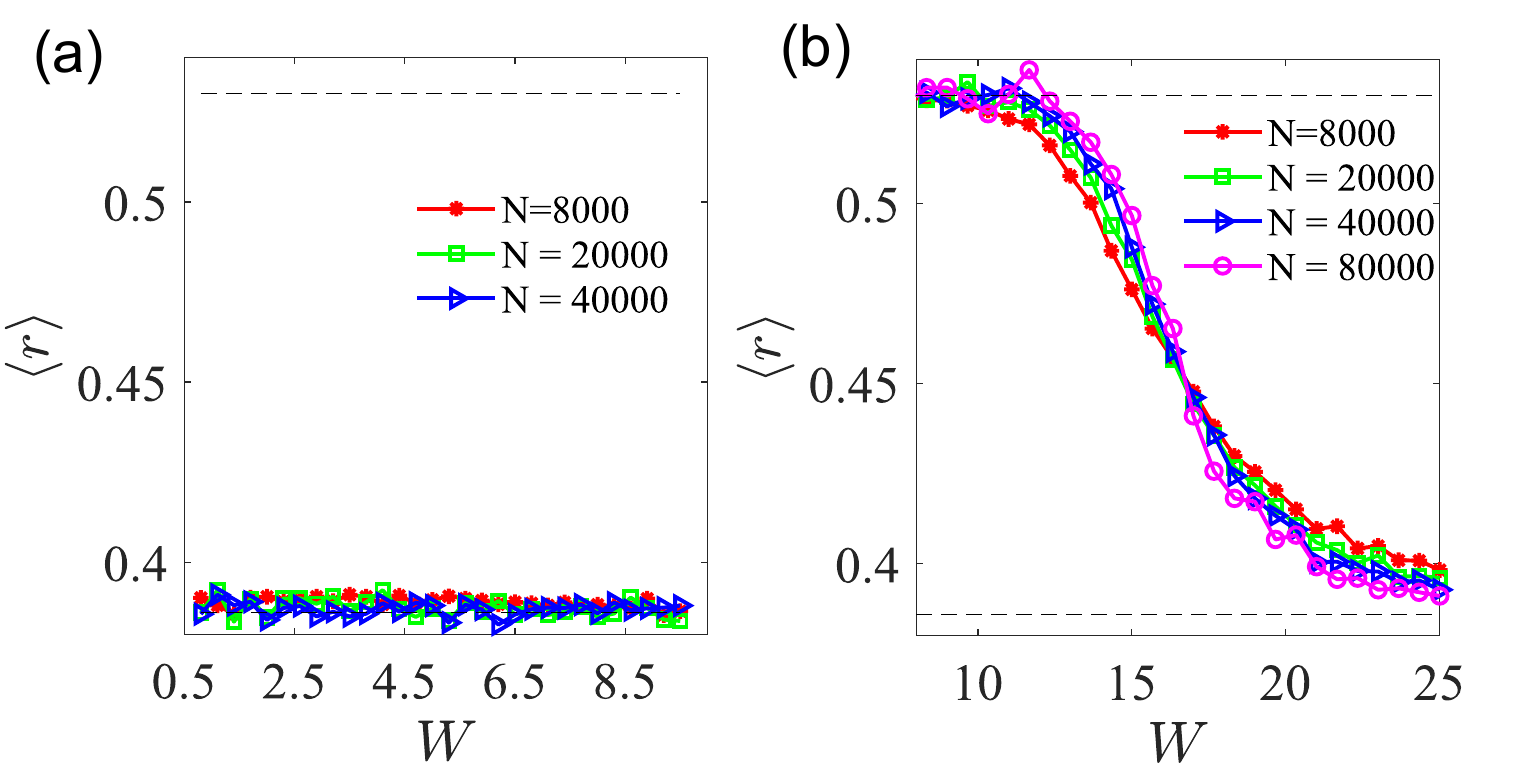}
\caption{Level statistics for (a) $(\eta,p) = (0,0)$ and (b) $(\eta,p) = (0,1)$. The average ratio of level spacings $\langle r\rangle$ is shown as a function of disorder strength $W$ for different system sizes. The averages are taken over the central 10 eigenstates and $2000$--$10000$ disorder realizations. The horizontal dashed lines indicate $\langle r\rangle = 0.387$ (Poisson) and $\langle r\rangle = 0.529$ (Wigner--Dyson).}
\label{figS1}
\end{figure}

\section{(III) Mean Fractal Dimension as a Function of Disorder Strength}

The mean fractal dimension $\langle D_2 \rangle$ as a function of disorder strength $W$ for $(\eta,p) = (0.2,1)$ and $(\eta,p) = (0,0.4)$ is shown in Fig.~S2(a) and (b), respectively. The averages are taken over the ten eigenstates closest to zero energy and over 4000 disorder realizations.

As shown in the figure, the curves for different system sizes $N$ intersect near the transition points, located around $W \approx 13$ in (a) and $W \approx 10$ in (b), respectively. These values are consistent with the transition points reported in the main text (Fig.~3), where the transition is identified from the level-statistics parameter $r$.

\begin{figure}[h]
\centering
\includegraphics[width=0.8\textwidth]{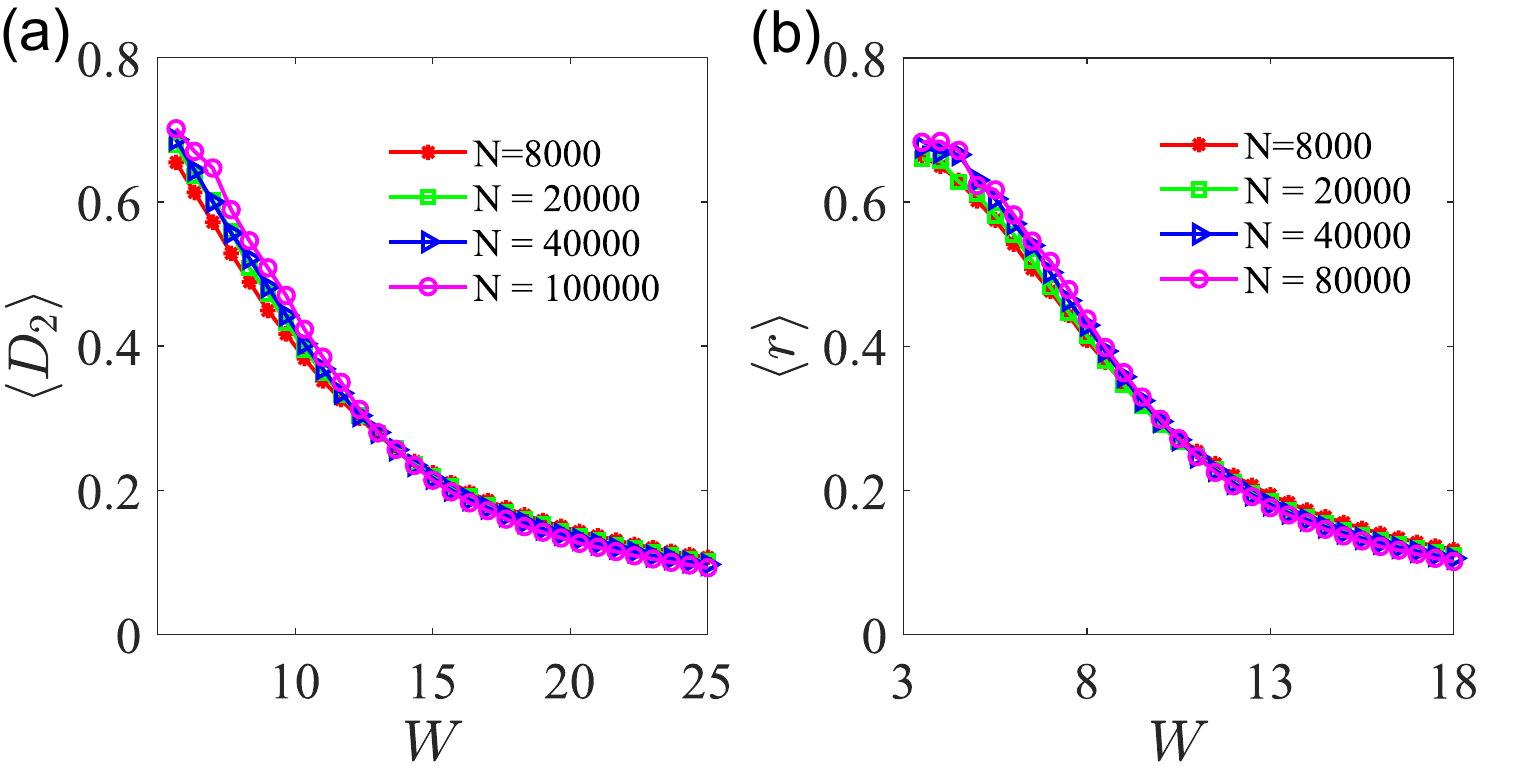}
\caption{Mean fractal dimension $\langle D_2 \rangle$ as a function of disorder strength $W$ for (a) $(\eta,p) = (0.2,1)$ and (b) $(\eta,p) = (0,0.4)$. Different curves correspond to different system sizes. The crossing points are located around $W \approx 13$(a) and $W \approx 10$(b), respectively, consistent with the transition points reported in the main text.}
\label{figS1}
\end{figure}

\end{document}